\documentclass[twocolumn,showpacs, prb]{revtex4}
\usepackage{graphicx}  
\usepackage{amssymb} 
\usepackage{amsmath}
\usepackage{enumerate}
\usepackage{textcomp}

\begin{document}
\title{Magnetic excitations in the spin-1 anisotropic antiferromagnet \textbf{NiCl$_2$-4SC(NH$_2$)$_2$}}
\author{C. Psaroudaki$^{1,2}$, S. A. Zvyagin$^{3}$, J. Krzystek$^{4}$, A. Paduan-Filho$^{5}$, X. Zotos$^{1,2,6}$, and N. Papanicolaou$^{1,6}$}
\affiliation{$^1$Department of Physics, University of Crete, 71003 Heraklion, Greece \\ $^2$Foundation for Research and Technology - Hellas, 71110 Heraklion, Greece\\
$^3$Dresden High Magnetic Field Laboratory (HLD),
Helmholtz-Zentrum Dresden-Rossendorf (HZDR),
01314 Dresden, Germany\\
$^4$National High Magnetic Field Laboratory, Florida State University, Tallahassee, FL 32310, USA\\
$^5$Instituto de Fisica, Universidade de Sao Paulo, 05315-970 Sao Paulo, Brazil\\
 $^6$Institute of Plasma Physics, University of Crete, 71003 Heraklion, Greece}
\date{\today}
\begin{abstract}
The spin-1 anisotropic antiferromagnet NiCl$_2$-4SC(NH$_2$)$_2$ exhibits a field-induced quantum phase transition that is formally analogous to Bose-Einstein condensation. 
Here we present results of systematic high-field electron spin resonance (ESR) experimental and theoretical studies of this compound with a special 
emphasis on single-ion two-magnon bound states. In order to clarify some remaining discrepancies between theory and experiment, the frequency-field 
dependence of magnetic excitations 
in this material is reanalyzed. In particular, a more comprehensive interpretation of the experimental 
signature of single-ion two-magnon bound states is shown to be fully consistent with theoretical results. We also clarify the structure of the ESR
 spectrum in the so-called intermediate phase. 
\end{abstract}

\maketitle
\section{Introduction}
\par
The organic compound NiCl$_2$-4SC(NH$_2$)$_2$ (known as DTN) is a gapped quasi-one-dimensional antiferromagnet with easy-plane anisotropy dominating the
 exchange coupling. At zero temperature, DTN undergoes a phase transition at a critical field $H_1\sim 2.1$ T above which nonzero spontaneous magnetization develops in the ground 
state and the spectrum of magnetic excitations becomes gapless. The system may then be thought of as a spin fluid formally described as a gas of hard-core bosons and the 
field-induced transition at $H_1$ corresponds to Bose-Einstein condensation. A further transition occurs at a second critical field
 $H_2\sim 12.6$ T above which the ground state is a fully ordered ferromagnetic state. The field-induced 
quantum phase transitions described above have already attracted considerable experimental interest through standard magnetometry, inelastic neutron scattering, specific 
heat measurements, etc., followed by theoretical calculations based on suitable Heisenberg models \cite{cite1}. 
\par
In particular, Zvyagin et al. \cite{cite2,cite3} have carried out detailed ESR measurements 
of magnetic excitations in a wide field range up to 25 T which includes the critical fields $H_1$ and $H_2$. The resulting picture was found
 to be generally consistent with early theoretical predictions \cite{cite4}. In short, in the low-field region $H<H_1$ the ground state carries 
zero azimuthal spin ($S_z=0$) while magnon excitations with $S_z=\pm1$ are separated by a gap which is unambiguously observed in the 
ESR spectrum through the expected $\Delta S_z=1$ transitions. In the high-field region, $H>H_2$, the ESR spectrum is dominated at 
low temperature by $\Delta S_z=1$ transitions between a fully ordered 
ferromagnetic ground state and magnons that again acquire a nonzero energy gap. At finite temperature, a new feature appears in the ESR 
spectrum and corresponds to a $\Delta S_z=1$ transition between a magnon and a single-ion two-magnon bound state, a fact anticipated theoretically 
sometime ago \cite{cite4}. A direct $\Delta S_z=2$ transition between the ordered ground state and a single-ion bound state is also observed.
 The physical picture becomes more involved in the intermediate field region $H_1<H<H_2$ but some progress has already 
been reported in recent literature \cite{cite5,cite6}. 
\par
Our current task is to clarify some remaining discrepancies between theory and experiment. 
Thus we have carried out afresh a new set of ESR experiments performed in a frequency range 
50-700 GHz using a tunable-frequency submillimeter-wave ESR spectrometer \cite{cite7} equipped with Backward Wave Oscillators as radiation sources and a 25 T resistive magnet.
A transmission-type probe with a sample in the  Faraday geometry was employed (with the light propagation 
vector directed along the applied magnetic field $H$ and the tetragonal c axis of the sample).
High-quality single crystals of DTN with a typical size of about $3\times3\times5$ mm$^3$ (from a new batch, grown from aqueous solutions of 
thiourea and nickel chloride) were used. 
A silicon-based Dow Corning High Vacuum Grease 976 V was used to fix samples inside the probe. Particular attention was paid to measuring the temperature dependence 
of the observed ESR modes, especially in order to unambiguously resolve the contribution of two-magnon bound states in the high-field region $H>H_2$. 
\par
The main body of the paper is devoted to a theoretical analysis carried out in two steps. In Sec.\ II we adopt a quasi-one-dimensional Heisenberg model to 
calculate important features of the ESR spectrum through a systematic strong-coupling expansion carried to third order. The general structure of the 
calculated spectrum agrees with experiment but important information concerning the relative intensity of the observed modes is practically impossible 
to obtain within this essentially three-dimensional (3D) model. Thus, in Sec. III, we carry out such calculations within a strictly one-dimensional 
(1D) model through exact diagonalization on finite chains and a corresponding simulation of the relevant dynamic susceptibilities. 
We are then able to analyze important features of the observed ESR spectrum over a wide field range including the intermediate region $H_1<H<H_2$. 
Our main conclusions are summarized in Sec. IV. 
\section{Three-dimensional Model}
\par
\begin{figure}
\centering
\includegraphics[scale=0.35]{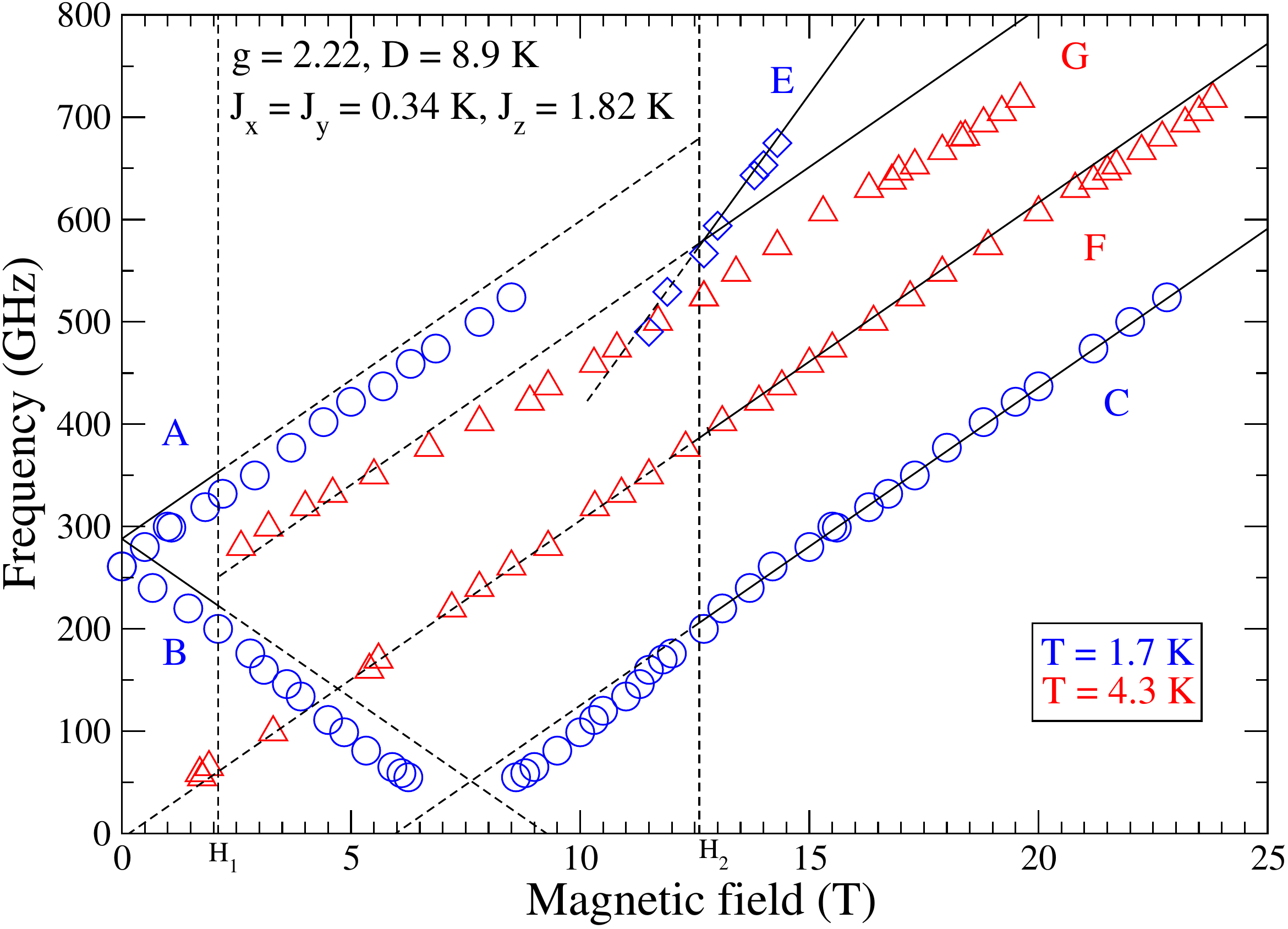}
\caption{(color online).\ Frequency-field dependence of magnetic excitations in DTN, with a uniform magnetic field $H$ applied along the tetragonal c axis.
 Blue symbols denote experimental data taken at T=1.7 K and red symbols at T=4.3 K. Note that the mode E was observed in the Voight configuration (with the light 
propagation vector directed perpendicular to the applied magnetic field) \cite{cite2,cite3} while the rest of the modes were observed in the Voight as well as in the Faraday 
geometry. Solid lines correspond to results of calculations presented in Sec. II 
and are deliberately continued as dashed lines into the intermediate region $H_1<H<H_2$. The location of critical fields $H_1=2.1$ T and $H_2=12.6$ T is indicated 
by vertical dashed lines.}\label{fig1}
\end{figure}

The essential features of the observed ESR spectrum are illustrated in Fig. \ref{fig1} together with some theoretical predictions 
derived from a spin $S=1$ Heisenberg Hamiltonian \cite{cite1,cite2}:
\begin{equation}
 \mathcal{H}=\sum_{i,\nu}J_\nu(\textbf{S}_i\cdotp \textbf{S}_{i+e_\nu}) +\sum_{i}[D(S_i^{z})^2+g\mu_BHS_i^z]
\label{eq1}
\end{equation}
\par
where $i$ denotes a generic site of a 3D lattice and $e_\nu$ with $\nu=\{x,y,z\}$ count nearest neighbors. The exchange constants $J_\nu=\{J_x,J_y,J_z\}$ 
may depend on the specific lattice direction and are assumed to be significantly smaller than the easy-plane anisotropy $(J_\nu<<D)$. 
Actually, DTN is thought to be described by the quasi-one-dimensional limit of Eq. (\ref{eq1}) defined from $J_x=J_y\ll J_z\ll D$, but the required theoretical 
analysis is essentially three-dimensional. Finally, an external magnetic field with strength $H$ is applied in a direction perpendicular to the easy-plane.
\par
At zero field ($H=0$) the ground state carries zero azimuthal spin $(S_z=0)$ and the magnon spectrum consists of two degenerate branches 
with $S_z=\pm1$ and energy-momentum dispersion $\omega=\omega(\bf{k})$ calculated through a systematic 1/$D$ expansion \cite{cite8} carried to third order:
\begin{eqnarray}
 \omega(\textbf{k})&=&D+2\sum_{\nu} J_{\nu}cosk_{\nu} \nonumber \\
&+&\frac{1}{D}\left[3 \sum_{\nu}J_{\nu}^{2}-2(\sum_\nu J_\nu cosk_\nu)^{2}\right]\nonumber \\
&+&\frac{1}{D^2}\left[2\sum_{\nu}J_{\nu}^3+4(\sum_\nu J_\nu cosk_\nu)^3 \right.\nonumber \\
&+&\frac{5}{2}\sum_\nu J_{\nu}^{3} cosk_\nu-7(\sum_{\mu}J_{\mu}^{2})(\sum_\nu J_\nu cosk_\nu)\nonumber \\
&-&\left. 2 (\sum_{\mu} J_{\mu}cosk_{\mu})(\sum_\nu J_\nu^2 cosk_\nu)\right]
\label{eq2}
\end{eqnarray}
\begin{figure}
\centering
\resizebox{\hsize}{!}{\includegraphics{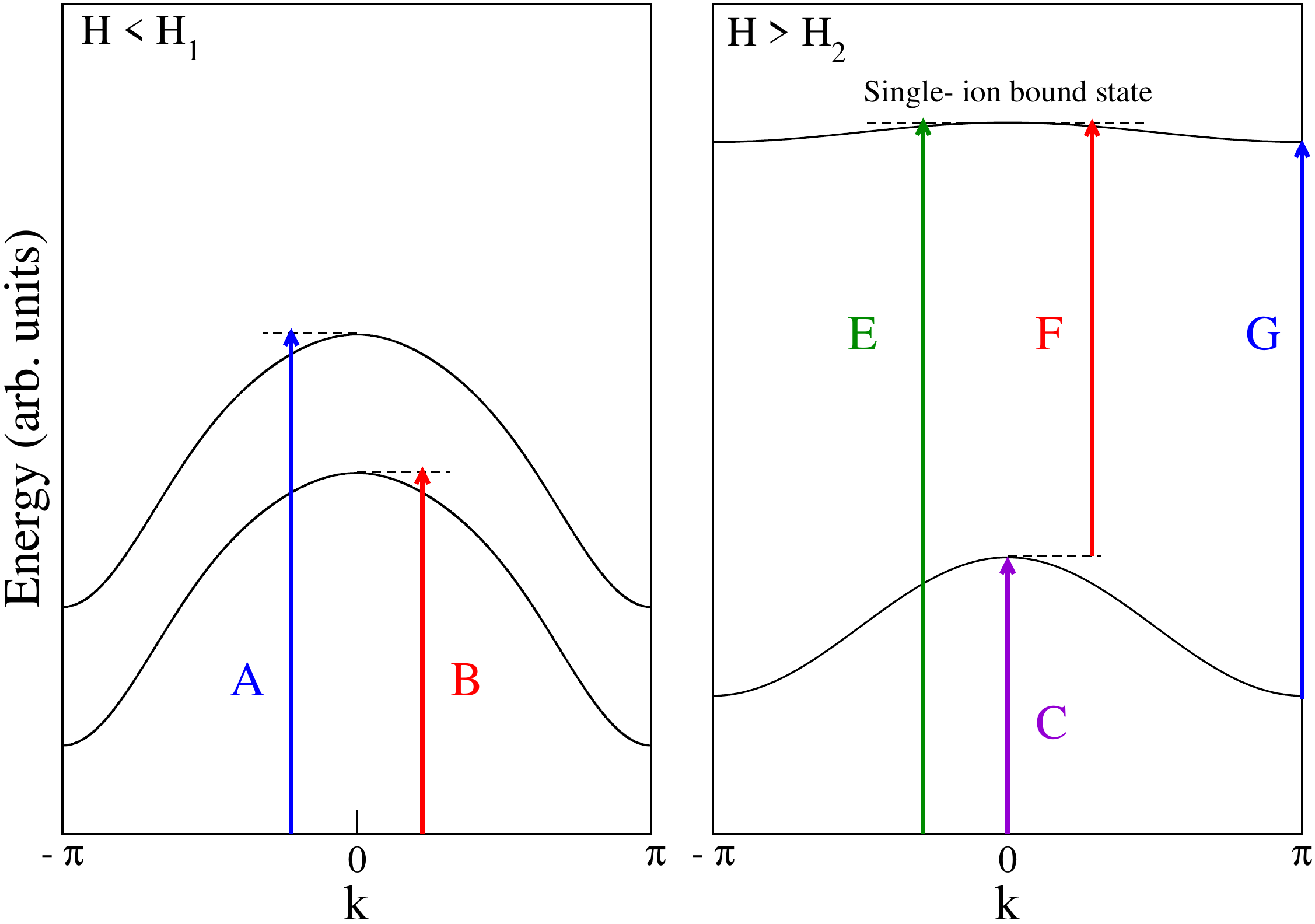}}
\caption{(color online). A schematic view of the energy-momentum dispersions of magnetic excitations in an $S=1$ Heisenberg chain with strong easy-plane ($D>0$) anisotropy for two typical fields $H<H_1$ (left) 
and $H>H_2$ (right). Note that the ESR transitions denoted by A, B, C, E and F occur at $k=0$, whereas transition G occurs at $k=\pi$. Two-particle continua are not shown for simplicity.}\label{fig1a}
\end{figure}
\par
For nonzero but sufficiently low fields the $S_z=0$ ground state remains unaffected while the degeneracy of the $S_z=\pm1$ magnon
 states is lifted (Fig. \ref{fig1a}, left) to yield a twofold dispersion:
\begin{equation}
 \omega_{\pm}(\textbf{k})=\omega(\textbf{k}) \pm g\mu_BH
\label{eq3}
\end{equation}
The ESR spectrum consists of two branches corresponding to $\Delta S_z=\pm1$ transitions between the ground state and $\textbf{k}=0$ magnons (modes A and B in Fig. \ref{fig1}). 
Thus the observed resonance frequencies are predicted to be:
\begin{equation}
 \omega_A=\omega_0+g\mu_BH ,\ \ \ \ \ \ \ \omega_B=\omega_0-g\mu_BH
\label{eq4}
\end{equation}
where $\omega_0=\omega(\textbf{k}=0)$ is calculated from Eq. (\ref{eq2}). Also note that the dispersion $\omega(\textbf{k})$ of Eq. (\ref{eq2})
 exhibits a nonzero gap throughout the Brillouin zone, the smallest gap occurring at $\textbf{k}=(\pi,\pi,\pi)$. 
Therefore, the magnon frequencies of Eq. (\ref{eq3}) remain positive throughout the zone as long as $H<H_1$ where $H_1$ is a critical field defined from:
\begin{equation}
 g\mu_BH_1=\Delta \ , \ \ \ \ \ \ \Delta=\omega[\textbf{k}=(\pi,\pi,\pi)] \ ,
\label{eq5}
\end{equation}
where the smallest gap $\Delta$ is again calculated from Eq. (\ref{eq2}) now applied for $\textbf{k}=(\pi,\pi,\pi)$. 
A corollary of the preceding discussion is that $\Delta<\omega_0$.
\par
When the field $H$ exceeds its critical value $H_1$ level crossing occurs and the azimuthal spin of the ground state no longer vanishes but increases 
with increasing field. Thus the system enters an intermediate phase through a field-induced quantum phase transition. The magnon spectrum 
is expected to be gapless in the intermediate phase but its detailed structure is now difficult to calculate. A systematic 1/$D$ expansion is not 
feasible while semiclassical methods are generally inaccurate at strong anisotropy. Hence we postpone further discussion of the intermediate phase
 until Sec. III where a numerical calculation is carried out within a strictly 1D model. 
\par
The theoretical model of Eq. (\ref{eq1}) becomes again tractable for sufficiently strong fields where the ground state is a completely 
ordered ferromagnetic state (Fig. \ref{fig1a}, right). 
The energy-momentum dispersion of single-magnon states is then given by 
\begin{eqnarray}
 \epsilon(\textbf{k})&=&g\mu_BH-D-2(J_x+J_y+J_z)\nonumber \\
&+& 2 (J_xcosk_x+J_ycosk_y+J_zcosk_z)
\label{eq6}
\end{eqnarray}
The lowest gap of this dispersion occurs at $\textbf{k}=(\pi,\pi,\pi)$ and is equal to $g\mu_BH-D-4(J_x+J_y+J_z)$. Thus the ordered state is stable when 
the field exceeds a critical value given by
\begin{equation}
 g\mu_BH_2=D+4(J_x+J_y+J_z)
\label{eq7}
\end{equation}
Therefore, for $H>H_2$, the ESR spectrum should be dominated by $\Delta S_z=1$ transitions between the completely ordered ferromagnetic state
and $\textbf{k}=0$ magnons. The resonance frequency is then calculated from 
\begin{equation}
 \omega_C=\epsilon(\textbf{k}=0)=g\mu_BH-D
\label{eq8}
\end{equation}
and is found to be independent of the exchange constants. 
\par
The physical picture is actually more involved for $H>H_2$ thanks to the appearance of an interesting class of two-magnon bound states. An exact 
calculation of such states is possible in the 1D model through an elementary Bethe Ansatz \cite{cite9}. The two-magnon spectrum contains 
a ``single-ion bound state'' whose energy-momentum dispersion extends well above the two-magnon continuum and was argued to be relevant for the analysis 
of the ESR spectrum observed in large-$D$ systems \cite{cite4}.
\par
However, a Bethe Ansatz is not applicable to the 3D model studied in this section. Thus we resort to a more direct method developed long time ago by Wortis
 \cite{cite10} for the calculation of two-magnon bound states in ferromagnets with arbitrary lattice dimension. The method is here generalized to account
for easy-plane anisotropy with strength $D$ and is employed in conjunction with the 1/$D$ expansion when analytical treatment is no longer feasible. Thus we were
 able to calculate the energy-momentum dispersion of the single-ion mode $E=E(\textbf{k})$ to third order in the 1/$D$ expansion. 
\par
We defer for the moment discussion of a $\textbf{k}=0$, $\Delta S_z=2$ transition between the ordered ground state and a single-ion bound state. 
Instead, we turn our attention to $\Delta S_z=1$ transitions between 
single magnons and single-ion bound states. These are absent at zero temperature but may occur with nonvanishing intensity at finite temperature. 
The corresponding resonance frequencies are then given by $\omega(\textbf{k})=E(\textbf{k})-\epsilon(\textbf{k})$ where $\textbf{k}$ extends over the entire 
Brillouin zone. Hence, at sufficiently low but nonzero temperature, resonance frequencies are expected to be observed throughout a band $\omega_F<\omega<\omega_G$ 
where the lower frequency is calculated to third order:
\begin{eqnarray}
 \omega_F&=&E(\textbf{k}=0)-\epsilon(\textbf{k}=0)\nonumber \\
&=&g\mu_BH+D-4(J_x+J_y+J_z)\nonumber \\
&+&\frac{2}{D}(J_x^2+J_y^2+J_z^2)+\frac{1}{D^2}(J_x^3+J_y^3+J_z^3)
\label{eq9}
\end{eqnarray}
whereas the upper frequency is given by
\begin{eqnarray}
 \omega_G&=&E[\textbf{k}=(\pi,\pi,\pi)]-\epsilon[\textbf{k}=(\pi,\pi,\pi)]\nonumber \\
&=&g\mu_BH+D
\label{eq10}
\end{eqnarray}
which is an exact result independent of the exchange constants, in analogy with the resonance frequency $\omega_C$ of Eq. (\ref{eq8}). 
\par 
\begin{figure}
\centering
\resizebox{\hsize}{!}{\includegraphics{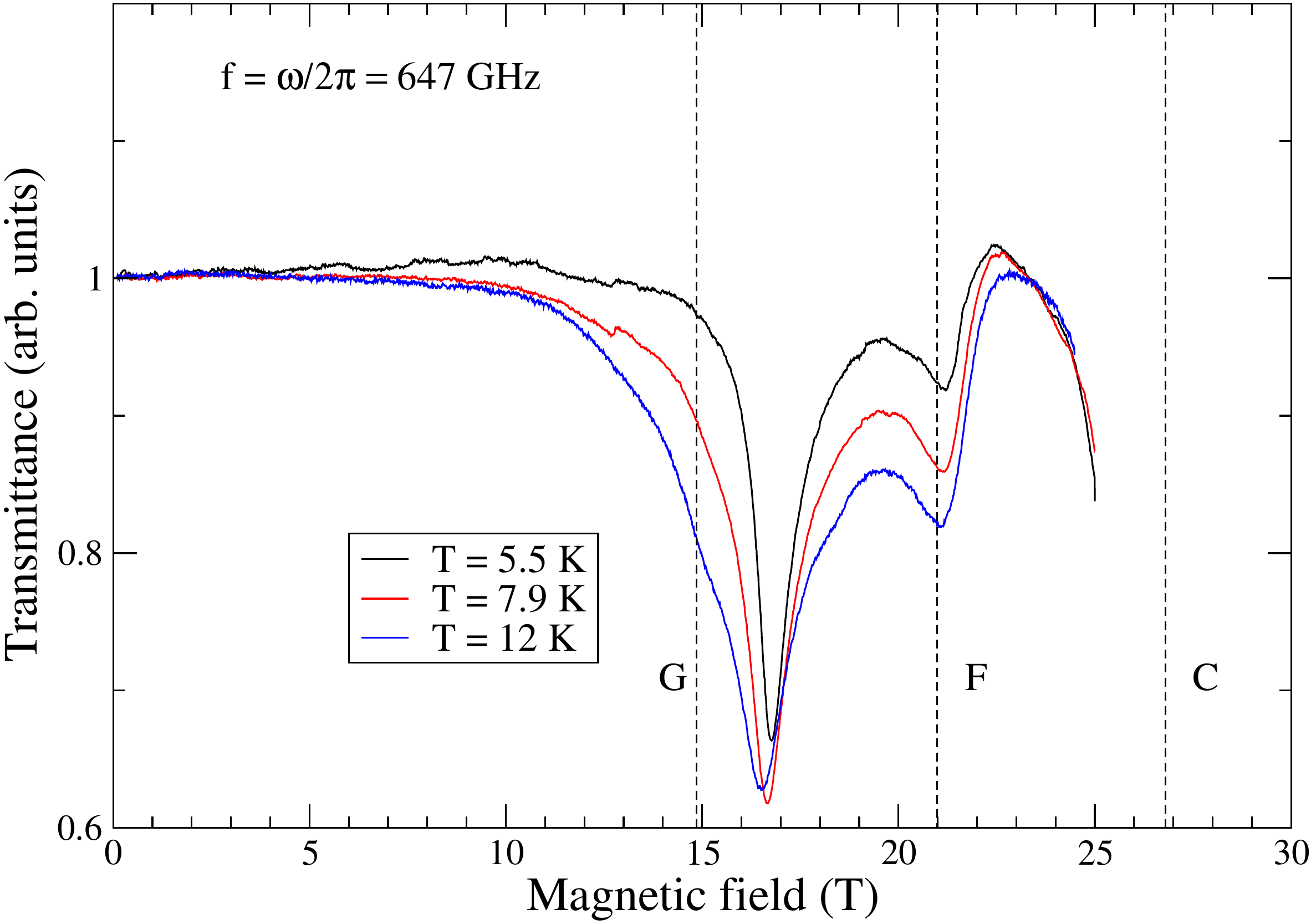}}
\caption{(color online). ESR transmittance spectra in DTN taken at frequency 647 GHz for three representative temperature values. 
Vertical line C indicates the location of the calculated single-magnon resonance, while F and G are the calculated boundaries of the single-ion (FG) band. 
Note that experiments were performed in magnetic fields up to 25 T and thus the single-magnon (C) resonance is not shown in this figure.}\label{fig2}
\end{figure}
\par
To summarize, the single-ion (FG) band is absent at zero temperature but acquires nonvanishing intensity at finite temperature. As was argued in Ref. \onlinecite{cite4} 
and further discussed in Sec. III of the present paper, the intensity is expected to display a characteristic double peak as a function of frequency 
at fixed external field, or as a function of field at fixed frequency. Therefore, both frequencies $\omega_F$ and $\omega_G$ are associated with the single-ion 
bound state and are relevant for the analysis of actual experiments. In this respect, it is worth mentioning that the absorption corresponding to mode G was initially observed 
in previous experiments \cite{cite3}. This absorption was interpreted as an artifact originating in the superficial layer of DTN crystals attacked 
by a GE-varnish solvent used to fix the sample within the sample holder. 
\par
\begin{figure}
\centering
\resizebox{\hsize}{!}{\includegraphics{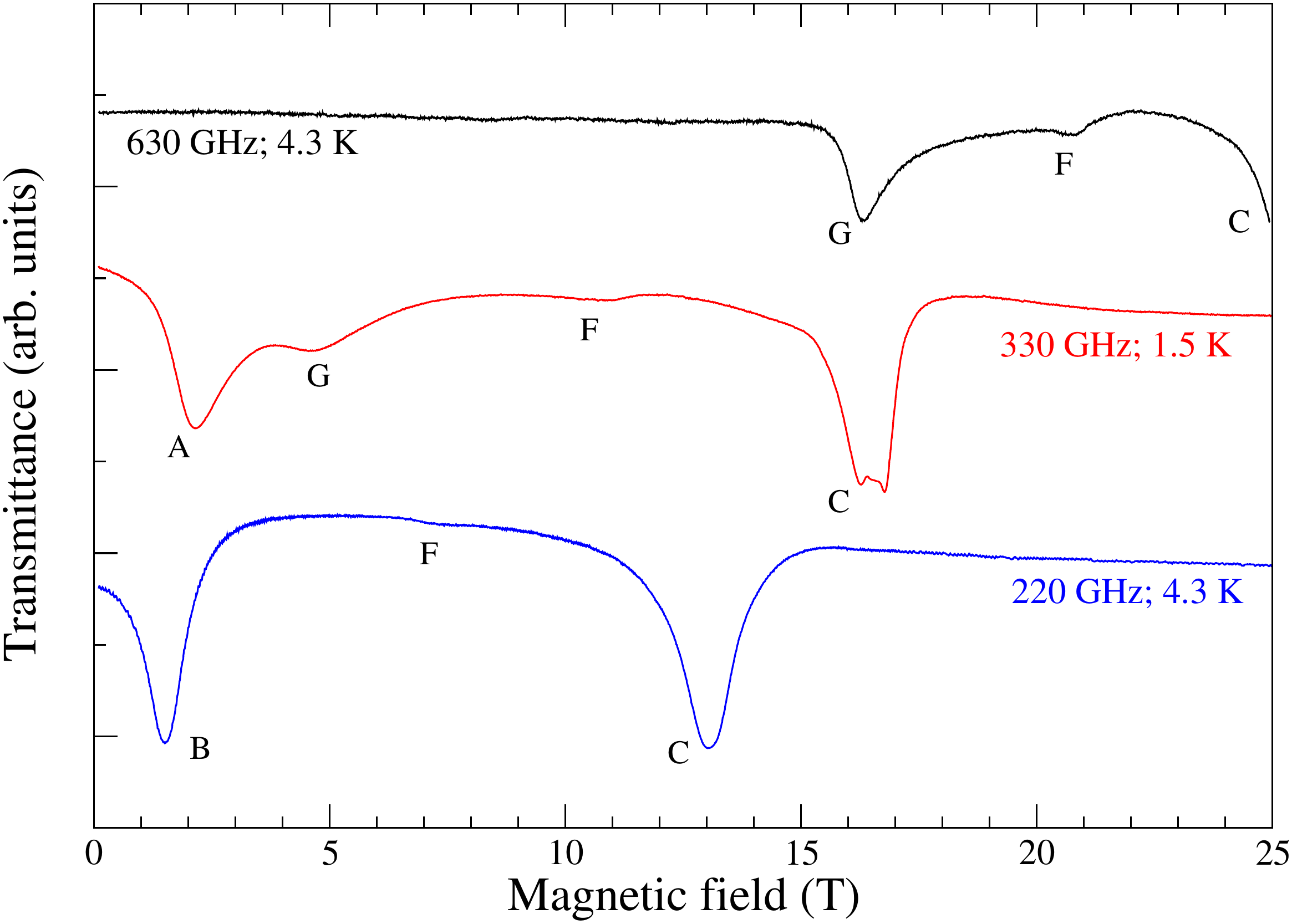}}
\caption{ (color online). ESR transmittance spectra for three characteristic frequencies and a wide field range up to 25 T. Note that the specific choice of frequencies 
is such that all possible modes appear in the figure.} \label{fig3a}
\end{figure}

However, our theoretical analysis suggests that the G mode is actually an inseparable partner in a doubly-peaked FG band associated with the single-ion 
bound state. Indeed, our current experiment supports such an interpretation, as shown in Fig. \ref{fig2} where the transmittance measured at fixed 
frequency $f=647$ GHz displays a characteristic double peak as a function of the applied field. Also note
that the double peak is uneven with most power absorbed for frequencies near the G boundary, an experimental fact that will be shown to be consistent 
with a numerical calculation of power absorption in Sec. III. Another important feature of Fig. \ref{fig2} is the apparent vanishing of intensity at relatively low temperatures 
(e.g., $T=1.5$ K), an experimental fact that is consistent with our interpretation of the FG resonance band as the result of transitions between excited states; namely, 
transitions between single magnons and single-ion two-magnon bound states (Fig. \ref{fig1a}, right). The overall picture suggested by Fig. \ref{fig2} is robust in the high-field 
region ($H>H_2$) and extends into the intermediate and low-field region ($H<H_2$) as shown in Fig. \ref{fig3a} and further discussed in Sec. III. 
\par
Finally, we return briefly to the possibility of a $\Delta S_z=2$ transition between the ordered ground state and a $\textbf{k}=0$ single-ion bound state, 
which would lead to a resonance frequency 
\begin{equation}
 \omega_E=E(\textbf{k}=0)=\omega_F+\omega_C
\label{eq11}
\end{equation}
but zero intensity thanks to the axial symmetry adopted in our theoretical models. However, crystal symmetry is compatible with some deviations from strict 
axial symmetry which apart from a tiny field misalignment may render mode E observable. In fact, such a mode was previously observed in DTN 
with a sample in the Voigt geometry \cite{cite2,cite3} and is included in Fig. \ref{fig1}. 
\par
\begin{figure}
\centering
\resizebox{\hsize}{!}{\includegraphics{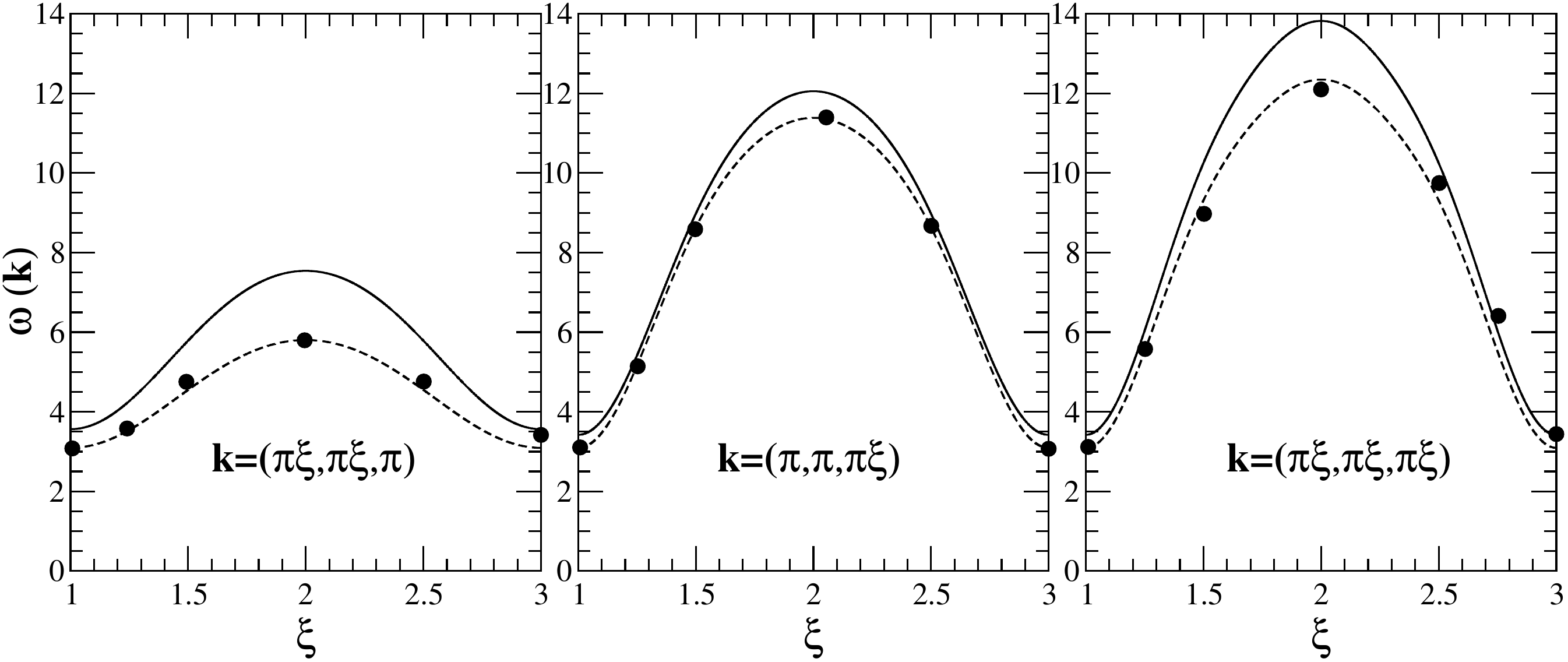}}
\caption{DTN dispersion of magnetic excitations at zero field calculated along three directions in the Brillouin zone using Eq. (\ref{eq2}) 
and parameters taken from Eq. (\ref{eq12})
(dashed lines) and Eqs. (\ref{eq13})-(\ref{eq14}) (solid lines). Symbols denote inelastic neutron scattering data taken from Ref. \onlinecite{cite1}. 
Energy is measured in degrees K.}\label{fig3}
\end{figure}

The remainder of this section is devoted to a brief discussion concerning the choice of suitable parameters. The simplest possibility is to fit the zero-field 
magnon dispersion given by Eq. (\ref{eq2}) to the dispersion measured via inelastic neutron scattering \cite{cite1}. A good fit is obtained, see Fig. \ref{fig3}, 
with the choice of parameters \cite{cite8}
\begin{equation}
 D=7.72\ \text{K},\ \ \ \ J_x=J_y=0.2\ \text{K}, \ \ \ \ J_z=1.86\ \text{K}  
\label{eq12}
\end{equation}
In particular, $\omega_0=\omega(\textbf{k}=0)=257$ GHz, in fair agreement with the experimental value $\omega_0=267$ GHz independently obtained through ESR. 
Consequently, the theoretically predicted branches A and B in the ESR spectrum agree with experiment if we further choose a gyromagnetic ration $g=2.22$. 
The critical field calculated from Eq. (\ref{eq5}) is $H_1=2.08$ T, in excellent agreement with the experimental value $H_1=2.1$ T. 
\par 
Thus the preceding choice of parameters yields a sufficiently accurate description of the low-field region $H<H_1$. But such a choice leads to poor 
quantitative predictions in the high-field region $H>H_2$. For example, the critical field $H_2$ calculated from Eq. (\ref{eq7}) is $H_2=11.24$ T, to be 
compared with the experimental $H_2=12.6$ T. Similarly, the exact magnon branch $\omega_C$ of Eq. (\ref{eq8}) substantially disagrees with experiment 
when $D=7.72$ K. 
\par 
Instead, an excellent fit of mode C is obtained using the parameters \cite{cite2} 
\begin{equation}
 g=2.22, \ \ \ \ D=8.9\ \text{K}
\label{eq13}
\end{equation}
We adopt these values and fix the remaining (exchange) constants via a least-square fit of the zero-field dispersion of Eq. (\ref{eq2}) to the experimental 
dispersion, see Fig. \ref{fig3}, to obtain 
\begin{equation}
 J_x=J_y=0.34\ \text{K}, \ \ \ \ J_z=1.82\ \text{K}
\label{eq14}
\end{equation}
which are significantly different from $J_x=J_y=0.18$ K and $J_z=2.2$ K obtained in Ref. \onlinecite{cite2} using a self-consistent semiclassical method to calculate
 the zero-field magnon dispersion. Here, to be consistent, we employ the parameters of Eq. (\ref{eq13}) and Eq. (\ref{eq14}) to calculate the critical fields
 $H_1=2.30$ T and $H_2=12.68$ T which are in rough agreement with the experimental $H_1=2.1$ T and $H_2=12.6$ T. We also employ the same parameters
 to calculate the branches of the ESR spectrum shown by straight lines A, B, C, F, G, E in Fig. \ref{fig1}, which are again in rough agreement with experiment. A schematic representation of the ESR 
transitions that correspond to the above modes is given in Fig. \ref{fig1a}. 
\par
\par
In our opinion, a completely satisfactory quantitative agreement with experiment should not be expected for two reasons. First, crystal symmetry of DTN allows, 
in principle, inclusion of several more terms of unknown strength in the effective Heisenberg Hamiltonian of Eq. (\ref{eq1}) such as Dzyaloshinskii-Moriya 
interactions between neighboring Ni sites \cite{cite5}. Second, even within the limits of model of Eq. (\ref{eq1}), 
accurate quantitative predictions may be difficult to obtain within the limited third-order 1/$D$ expansion, while similar uncertainties may
 arise in the context of sophisticated semiclassical methods \cite{cite2}. Nevertheless, we believe that the overall qualitative picture is substantially correct. 
\section{One-dimensional Model}
\par
Yet there are several questions that are difficult to settle within the essentially 3D model of Eq. (\ref{eq1}), such as the calculation of intensities of the
 various ESR modes, the structure of the spectrum in the intermediate phase, etc. But the essential features of the observed spectrum are already 
accounted for by a relatively simple 1D model which is further employed in the present paper to discuss the remaining questions. 
\par
In order to establish consistency with the earlier work \cite{cite4} we first restrict the main results of Sec. II to the 1D model through the formal substitution
\begin{equation}
 J_x=J_y=0,\ \ \ \ J_z=J
\label{eq15}
\end{equation}
For instance, the Heisenberg Hamiltonian of Eq. (\ref{eq1}) now reads
\begin{equation}
 \mathcal{H}=\sum_{n=1}^N [J(\textbf{S}_n\cdot \textbf{S}_{n+1})+D(S_n^z)^2+g\mu_BHS_n^z]
\label{eq16}
\end{equation}
where the summation extends over the N sites of a 1D lattice assumed to be periodic. 
\par
The zero-field magnon dispersion of Eq. (\ref{eq2}) reduces to 
\begin{eqnarray}
\omega(k)&=& D \left[ 1+\frac{2 J}{D}cosk+\frac{J^2}{D^2}(1+2 sin^2k)\right. \nonumber \\
&+& \left. \frac{J^3}{D^3}[2sin^2k-\frac{1}{2}(1+8 sin^2k)cosk]\right]
\label{eq17}
\end{eqnarray}
and agrees with an early calculation within the 1D model \cite{cite11}. Recently, several more terms beyond the third order have become available
 \cite{cite12,cite13}. However, for anisotropy strengths of current interest, the third-order result (\ref{eq17}) proves to be sufficiently accurate.
\par
In particular, the ESR modes A and B may be calculated from Eq. (\ref{eq4}) now applied with 
\begin{equation}
 \omega_0=\omega(k=0)=D\left(1+\frac{2 J}{D}+\frac{J^2}{D^2}-\frac{J^3}{2 D^3}\right)
\label{eq18}
\end{equation}
Similarly, the critical field $H_1$ is given by Eq. (\ref{eq5}) applied with
\begin{equation}
 \Delta=\omega(k=\pi)=D\left(1-\frac{2 J}{D}+\frac{J^2}{D^2}+\frac{J^3}{2 D^3}\right)
\label{eq19}
\end{equation}
Analogous results may be obtained for sufficiently strong fields in the region $H>H_2$ where the ground state is a fully ordered ferromagnetic state. 
The single-magnon dispersion of Eq. (\ref{eq6}) reduces to
\begin{equation}
 \epsilon(k)=g\mu_BH-D-2 J(1-cosk)
\label{eq20}
\end{equation}
whose lowest gap occurs at $k=\pi$ and is equal to $g\mu_BH-D-4J$. Therefore, the upper critical field is given by
\begin{equation}
 g\mu_BH_2=D+4J,
\label{eq21}
\end{equation}
which agrees with the 1D reduction of Eq. (\ref{eq7}). For $H>H_2$, the domimant resonance frequency $\omega_C$ arises from $\Delta S_z=1$ 
transitions between the ordered state and $k=0$ magnons:
\begin{equation}
 \omega_C=\epsilon(k=0)=g\mu_BH-D
\label{eq22}
\end{equation}
which is exact and coincides with the 3D result of Eq. (\ref{eq8}).
\par
The dispersion of the single-ion two-magnon bound state can be calculated exactly within the 1D model \cite{cite4,cite9} but a third-order approximation 
is sufficient for our purposes:
\begin{equation}
 E(k)=2 g\mu_BH+D\left[-\frac{4J}{D}+(\frac{2J^2}{D^2}+\frac{J^3}{D^3})cos^2\frac{k}{2}\right]
\label{eq23}
\end{equation}
Thus a $\Delta S_z=1$ transition between a $k=0$ magnon and a $k=0$ single-ion two-magnon bound state yields a resonance frequency
\begin{eqnarray}
 \omega_F&=&E(k=0)-\epsilon(k=0)\nonumber \\
&=&g\mu_BH+D\left[1-\frac{4J}{D}+\frac{2J^2}{D^2}+\frac{J^3}{D^3}\right]
\label{eq24}
\end{eqnarray}
which agrees with the 1D reduction of Eq. (\ref{eq9}), whereas a transition between a $k=\pi$ magnon and a $k=\pi$ single-ion bound state yields
\begin{equation}
 \omega_G=E(k=\pi)-\epsilon(k=\pi)=g\mu_BH+D
\label{eq25}
\end{equation}
which is exact (independent of $J$) and coincides with the 3D result of Eq. (\ref{eq10}). Similar transitions occur for other values of k throughout the 
Brillouin zone and lead to a band of resonance frequencies in the region $\omega_F<\omega<\omega_G$. Although the intensity of such transitions vanishes at zero temperature, 
nonzero intensity is expected to occur at finite temperature, an issue to be discussed in detail in the continuation of this section.
\par
First, a digression concerning the choice of parameters within the 1D model. Recall that the single-magnon resonance frequency $\omega_C$ given in 
Eq. (\ref{eq22}) is an exact prediction of the 1D as well as the 3D model; see Eq. (\ref{eq8}). Therefore, we adopt in this section the choice of 
the gyromagnetic ratio $g$ and anisotropy $D$ already made in Eq. (\ref{eq13}) and the only remaining parameter is the exchange constant $J$ or, equivalently, 
the dimensionless ratio $J/D$. A semi-quantitative agreement with experiment is obtained with the choice
\begin{equation}
 g=2.22,\ \ \ \ D=8.9\ \text{K},\ \ \ \ \frac{J}{D}=\frac{1}{4}
\label{eq26}
\end{equation}
which will be adopted in all calculations presented in this section. For convenience, we use rationalized variables such that frequency
 $f=\omega/2 \pi$ is measured in units of $D/2 \pi \hbar=185.45$ GHz, magnetic field $h=g\mu_BH/D$ in units of $D/g\mu_B=6$ T and temperature $\tau=T/D$ 
in units of $D=8.9$ K. 
\par
The main issue addressed in this section is an explicit calculation of power absorption. In a typical ESR experiment a microwave field of angular frequency 
$\omega$ is applied in the basal plane along, say, the x-axis, in addition to a uniform bias field $H$ applied along the z-axis. The intensity or power 
absorption per site is defined up to an overall multiplicative constant by
\begin{equation}
 I \sim \omega  \chi '' (\omega)/N
\label{eq27}
\end{equation}
where the imaginary part of the susceptibility is given by \cite{cite14}
\begin{equation}
 \chi''(\omega)=\frac{\pi}{Z}\sum_{a,b}(e^{-\beta E_b}-e^{-\beta E_a})|\langle a|\mu_{x}|b|\rangle|^2\delta(E_a-E_b-\omega)
\label{eq28}
\end{equation}
Here sums extend over all eigenstates $|a\rangle$ of Hamiltonian of Eq. (\ref{eq16}), $E_a$ are the corresponding eigenvalues, $\beta=1/T$ is the inverse temperature, 
and $Z=\sum_a e^{-\beta E_a}$ is the total partition function. Finally, matrix elements in Eq. (\ref{eq28}) involve the total spin operator in the x-direction
 $\mu_x=\sum_n S_n^x$. 
\par
An analytical calculation of $\chi''(\omega)$ is out of question except in very special limits \cite{cite4}. We thus resort to a numerical calculation based 
on Eq. (\ref{eq28}) and a complete diagonalization of the spin-1 Hamiltonian of Eq. (\ref{eq16}) defined on a finite periodic chains with size N as large as 12. 
Actually, explicit results presented below were obtained on a chain with N=10, whereas N=12 chains were occasionally used for consistency checks. 
On a finite chain Eq. (\ref{eq28}) yields a susceptibility that is a sum of weighted $\delta$-functions and is thus rather spiky. 
Hence we adopted an empirical smoothing process to obtain an intensity 
\begin{equation}
 I=I(f,h,\tau)
\label{eq29}
\end{equation}
that is a reasonably smooth function of frequency $f$, magnetic field $h$, and temperature $\tau$, measured in rationalized physical units defined in the text
 following Eq. (\ref{eq26}). Our main task is then to analyze the calculated intensity as a function of all three variables. 
\begin{figure}
\centering
\resizebox{\hsize}{!}{\includegraphics{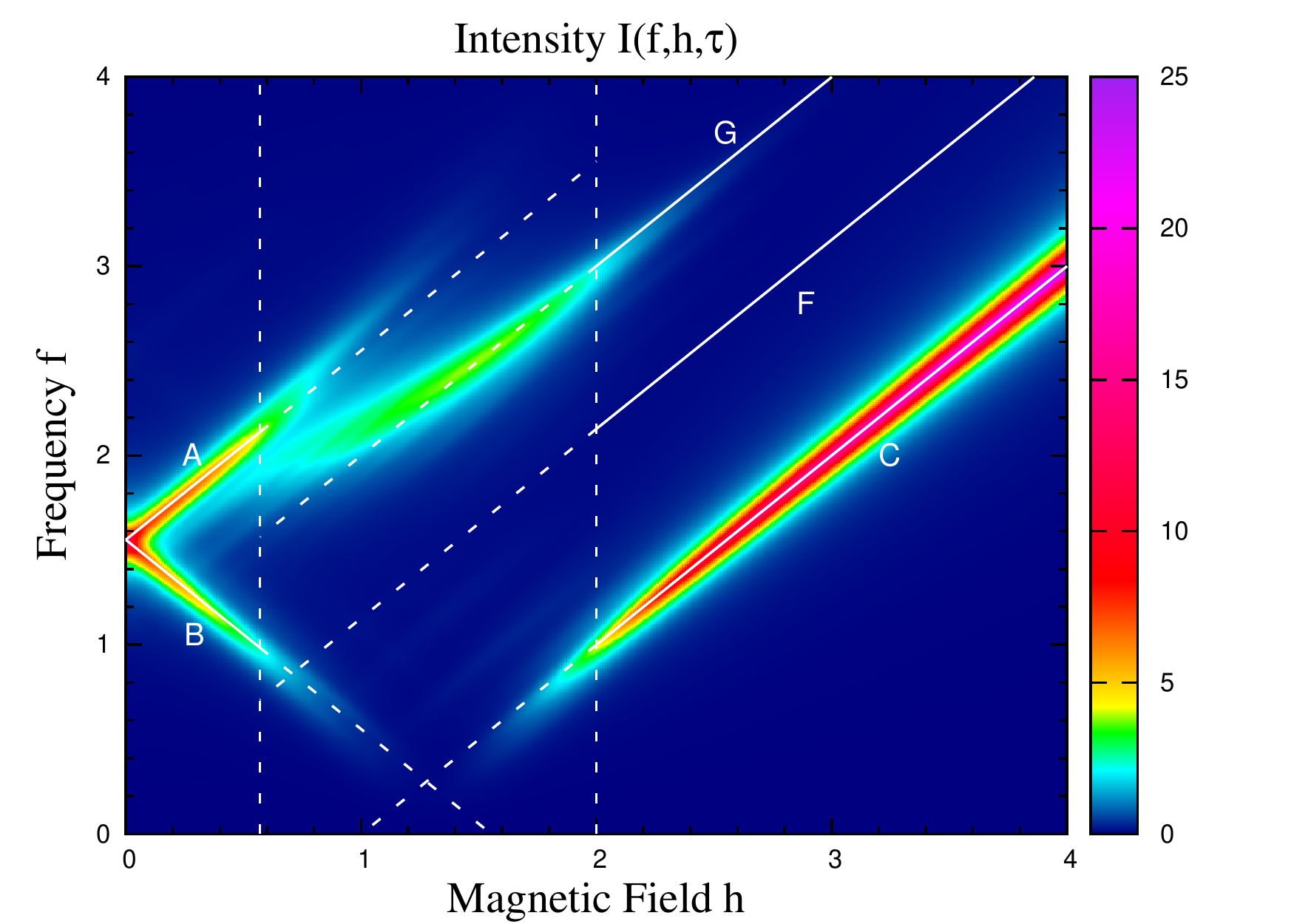}}
\caption{(color online).\ Colored surface represents the normalized ESR intensity $I(f,h,\tau)$ calculated for a spin chain with N=10 as a function 
of frequency $f$ and magnetic field $h$, at fixed temperature $\tau=0.2$ ($T=1.8$ K). Solid lines correspond to results of calculation within the 1D model 
of Sec. III and are deliberately extended as dashed lines into the intermediate region $h_1<h<h_2$. The location of critical fields $h_1=0.57$ and 
$h_2=2$ is indicated by vertical dashed lines. Note that $f$, $h$ and $\tau$ are all measured in rationalized units defined after Eq. (\ref{eq26}). }\label{fig4}
\end{figure}
\begin{figure}
\centering
\resizebox{\hsize}{!}{\includegraphics{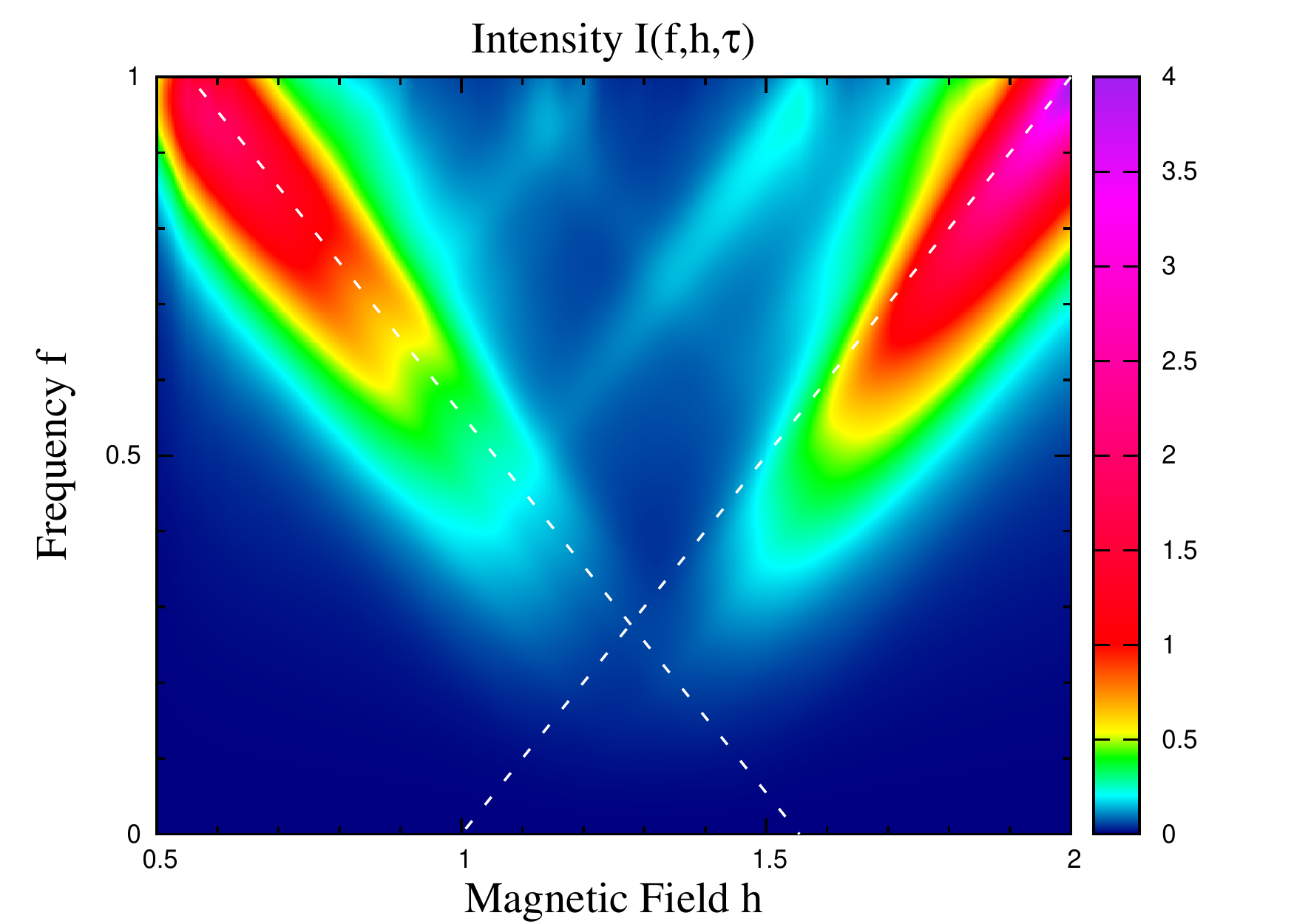}}
\caption{(color online).\ Same as Fig. \ref{fig4} but now focusing on the low-frequency end of the intermediate phase $h_1<h<h_2$. Note the formation of a V-like 
structure with rapidly decreasing intensity as one approaches the center of the intermediate phase. }\label{fig5}
\end{figure}
\par
\begin{figure}
\centering
\resizebox{\hsize}{!}{\includegraphics{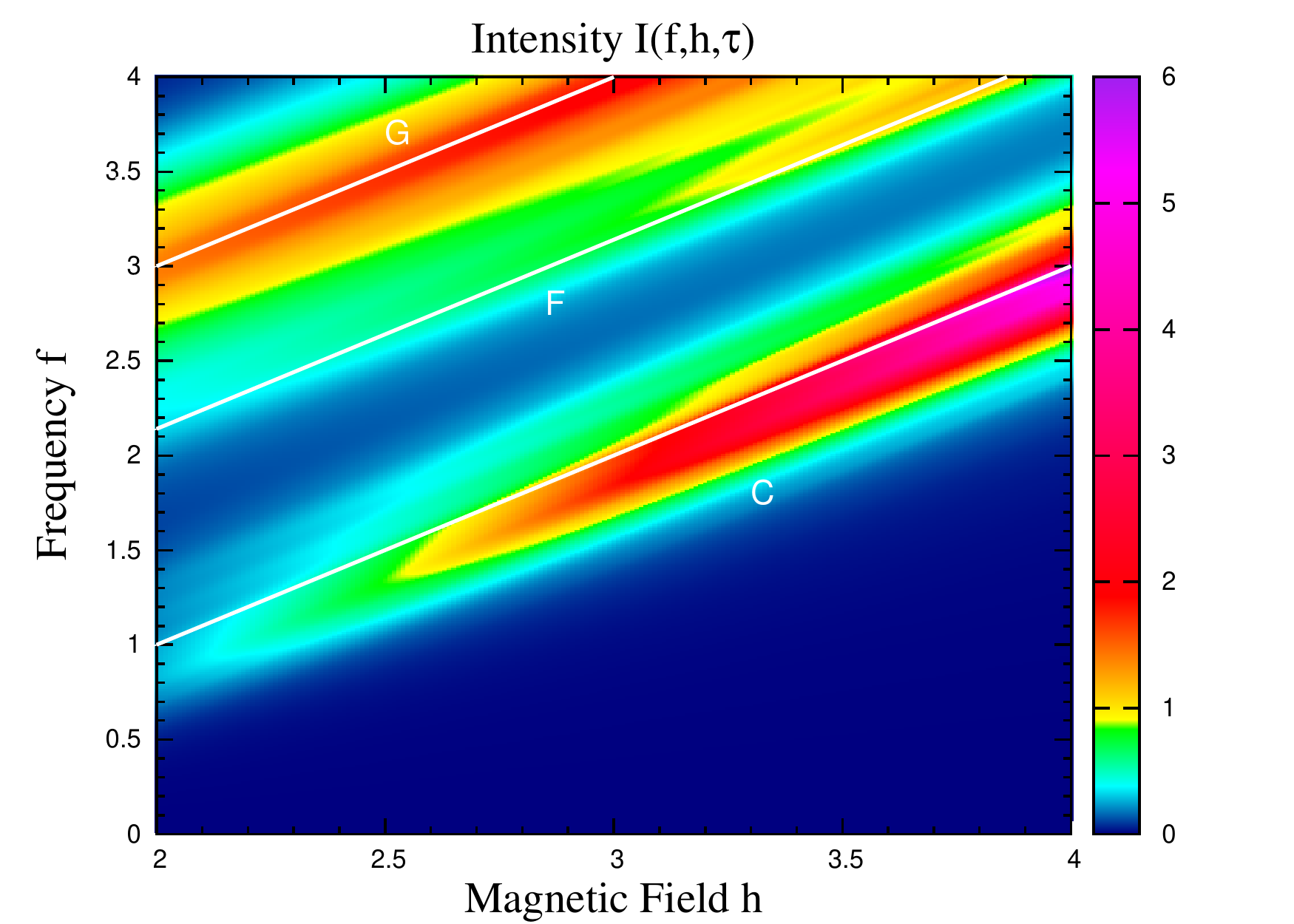}}
\caption{(color online).\ Same as Fig. \ref{fig4} but with intensity now calculated at a much higher temperature $\tau=2$ $(T=18\ K)$. 
Note that we concentrate on the high-field region $h>h_2=2$ $(H>12\ T)$ in order to emphasize the significant enhancement of intensity in the FG band 
(especially near the G boundary) which provides unambiguous evidence for the existence of a single-ion two-magnon bound state. }\label{fig6}
\end{figure}

In Fig. \ref{fig4} we present our results for the (colored) surface $I=I(f,h)$ at fixed temperature $\tau=0.2$ ($T=1.8$ K) which is a typical relatively 
low temperature of experimental interest \cite{cite2}. Superimposed in the same figure are the analytical predictions for the two critical fields $h_1=0.57$ 
and $h_2=2$ calculated from Eq. (\ref{eq19}) and Eq. (\ref{eq21}) adapted to rationalized units, as well as corresponding predictions for the resonance lines 
A, B for $h<h_1$ and C, F, G for $h>h_2$ calculated earlier in this section. Note that mode E (see Fig. \ref{fig1}) does not appear in Fig. \ref{fig4} because it 
corresponds to a $\Delta S_z=2$ transition 
and its intensity $I$ vanishes within the strictly axially symmetric model of Eq. (\ref{eq16}). Several important facts have already become apparent in 
Fig. \ref{fig4} which we analyse in turn:
 \begin{enumerate}[(a)]
\item We note that the magnon resonance lines A, B and C roughly coincide with the maxima of the calculated intensity as expected in the low-temperature 
region. Nevertheless, the chosen temperature $T=1.8$ K is sufficiently high to account for the anticipated line broadening which is also apparent in Fig. \ref{fig4}. 
Yet, this temperature is too low to yield a significant signal for the single-ion bound state, as shown in Fig. \ref{fig4} where the intensity practically vanishes 
in the FG region. 
\item We examine the results of Fig. \ref{fig4} in the intermediate region $h_1<h<h_2$ where analytical predictions are practically absent. The most conspicuous feature
 is a tail of line G with strong intensity in the intermediate region, which persists even at very low temperature where line G itself looses its intensity 
for $h>h_2$. Therefore, the G-tail corresponds to some sort of a collective excitation that is robustly present in our current experiment and requires further 
theoretical investigation. On the other hand, mode F acquires a tail into the intermediate region with intensity that diminishes at low temperature and is thus invisible in 
Fig. \ref{fig4}.
\item We note that the magnon lines A, B and C also acquire tails but with intensity that gradually vanishes as one approaches the center of the 
intermediate phase. The structure of the tails becomes apparent in Fig. \ref{fig5} which focuses on the low-frequency end of the intermediate phase. Thus we reveal 
a V-like structure with intensity that gradually vanishes as one approaches the center. This picture apparently contradicts the result of Cox et al.
 \cite{cite6} who predict by a similar calculation a Y-like structure with intensity that remains finite and practically constant near the center. 
On the other hand, our result is consistent with a rounding of a V into a U structure predicted to occur in the presence of a small Dzyaloshinskii-Moriya 
anisotropy treated by a semiclassical method \cite{cite5}.
\item As mentioned already, Fig. \ref{fig4} as well as experiment indicate absence of measurable intensity in the single-ion (FG) band at the relatively low temperature 
$\tau=0.2$ ($T=1.8$ K). However, the FG band is significantly activated at higher temperature, as demonstrated in Fig. \ref{fig6} which displays the intensity
 in the field region $h>h_2=2$ ($H>12$ K) and temperature $\tau=2$ ($T=18$ K). The calculated FG band is highly populated at this temperature with most 
of the intensity concentrated near the G boundary. 
\end{enumerate}
\par
\begin{figure}
\centering
\resizebox{\hsize}{!}{\includegraphics{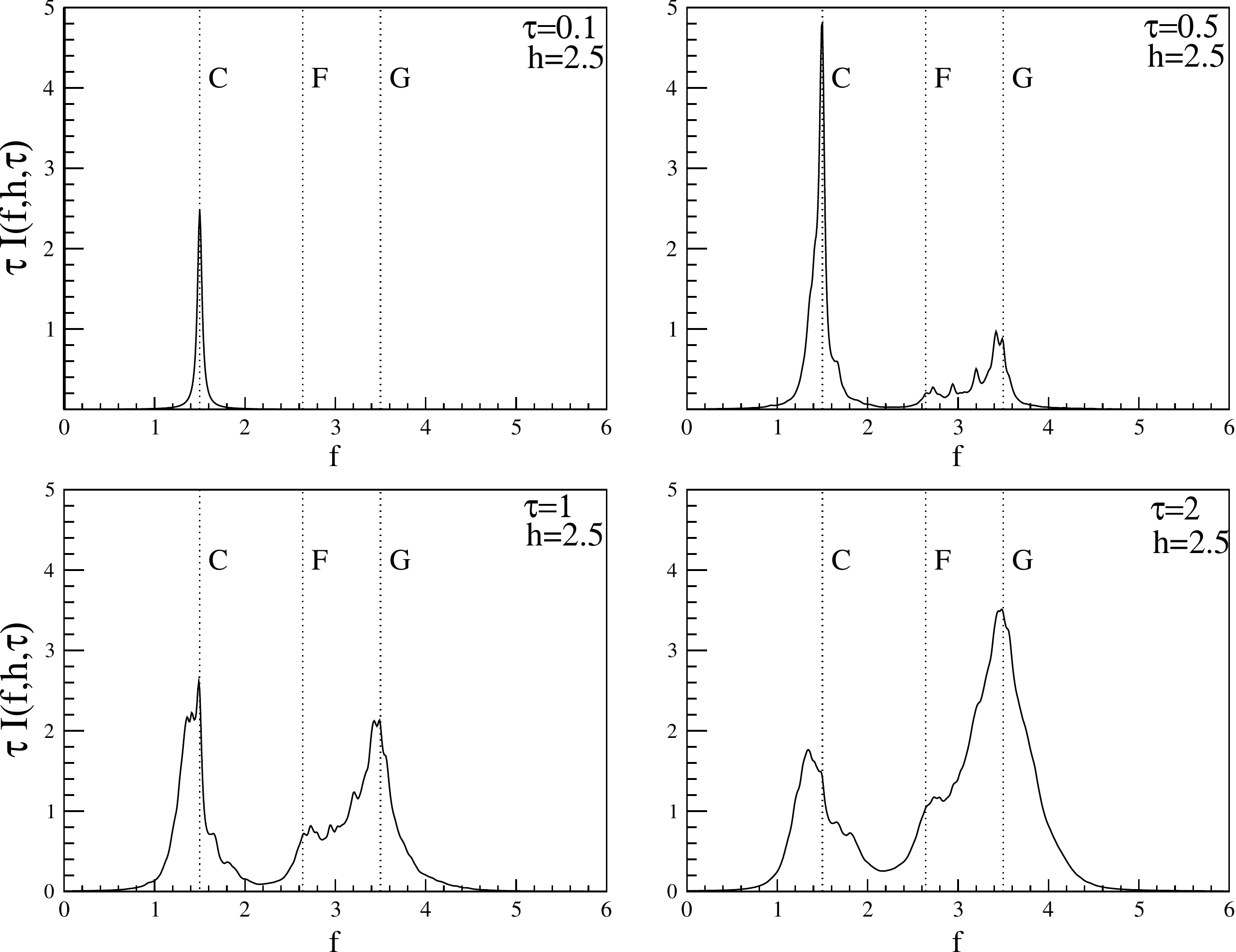}}
\caption{Calculated normalized intensity $I(f,h,\tau)$ scaled with temperature $\tau$, as a function of frequency $f$ at a fixed field $h=2.5$ and four 
typical values of temperature $\tau$. Vertical lines C, F and G indicate the location of the single-magnon resonance C and the boundaries of the 
single-ion (FG) band calculated within the 1D model. Recall that $f$,$h$ and $\tau$ are all measured in rationalized units defined after Eq. (\ref{eq26}).}\label{fig7}
\end{figure}

To understand the preceding result in some detail, we depict in Fig. \ref{fig7} the intensity as a function of frequency at a fixed field $h=2.5$ ($H=15$ T) and selected 
values of temperature. At the lowest temperature $\tau=0.1$ ($T=0.9$ K) considered in Fig. \ref{fig7}, the dominant feature is the magnon resonance C while there is no 
sign for a single-ion bound state. On the contrary, the FG band is activated already at temperature $\tau=0.5$ ($T=4.5$ K) employed in actual experiments 
\cite{cite2,cite3}. The FG signal is further enhanced at higher temperature, as is evident in the $\tau=1$ ($T=9$ K) and $\tau=2$ ($T=18$ K) entries. 
Also evident is the formation of a double peak in the FG region, with the dominant peak occurring near the G boundary while a peak of lower intensity 
develops near the F boundary. The relative enhancement of the intensity near the G boundary is likely due to the fact that it involves transitions between 
$k=\pi$ single magnons and $k=\pi$ single-ion bound states, where the magnon acquires its lowest gap and is thus more heavily populated at finite temperature than,
 say, $k=0$ magnons. 
\par
\begin{figure}
\centering
\resizebox{\hsize}{!}{\includegraphics{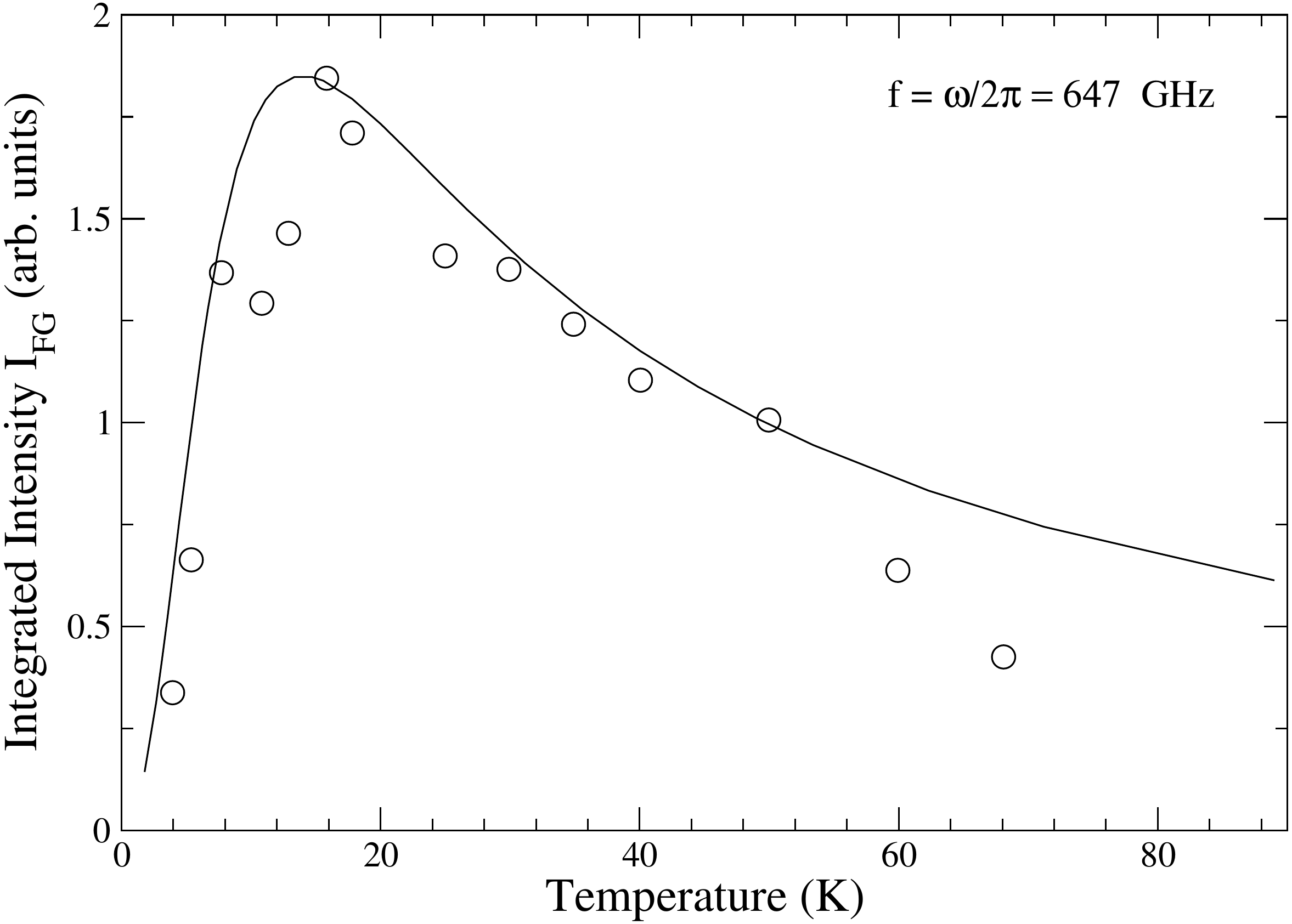}}\caption{Total (integrated) intensity of the FG band as a function of temperature at fixed frequency $f=647$ GHz calculated within the 1D model. 
Symbols denote experimental data extracted from field integration of ESR spectra.}\label{fig8}
\end{figure}

 The preceding theoretical findings are consistent with experimental results of the type shown in Fig. \ref{fig2}, with due attention to the fact that Fig. \ref{fig2} depicts the transmittance
 as a function of applied field at fixed frequency. In any case, both theory and experiment 
suggest a dominant peak near the G boundary followed by a secondary peak (a knee) near the F boundary. The two peaks are partners in a doubly-peaked FG band 
that cannot be separated in any meaningful way. Thus it is difficult to measure or calculate their relative intensity. Nevertheless, it is possible to 
calculate the total intensity of the FG band:
\begin{equation}
I_{FG}=\int_a^b I(f,h,\tau) dh\ \ , 
\label{eq30}
\end{equation}
where integration extends over a field interval $a\leq h \leq b$ chosen empirically so that it encompass the entire FG band. The total intensity of Eq. (\ref{eq30}) 
is depicted in Fig. \ref{fig8} as a function of temperature at fixed frequency $f=647$ GHz, together with experimental results obtained by applying a similar 
integration process to data of the type shown in Fig. \ref{fig2}. Taking into account that intensity is displayed in ``arbitrary units'', the qualitative agreement 
between theory and experiment shown in Fig. \ref{fig8} is satisfactory and fully consistent with our current interpretation of the ESR signal of the single-ion 
two-magnon bound state. 

\section{Conclusions}
\par
As far as the general structure of the observed ESR spectrum is concerned, the theoretical predictions of the 3D model of Eq. (\ref{eq1}) and the 1D model of Eq. (\ref{eq16})
are qualitatively similar and in reasonable quantitative agreement with experiment. But a detailed investigation of the remaining discrepancies required a 
calculation of the intensities of the various ESR modes, which is not feasible within the 3D model. Thus most of our effort was devoted to a detailed numerical 
calculation of intensity within the 1D model. The main new results are the following:
\par
While there have been numerous theoretical predictions for the occurrence of two-magnon bound states in quantum spin systems, experimental observation
 has been rather slow. Perhaps, the most interesting feature of the ESR spectrum in large-$D$ systems is the evidence it provides for the existence of the 
so-called single-ion two-magnon bound states. The original theoretical suggestion was made some time ago \cite{cite4} and was thought to explain ESR data 
obtained on a large-$D$ compound abbreviated as NENC \cite{cite15,cite16,cite18}. But a thorough experimental investigation was carried out more recently in 
relation to the title compound (DTN) \cite{cite2,cite3}.
\par
Our present investigation clearly suggests that the F and G lines are inseparable 
partners in a doubly-peaked FG band which originates in transitions between single magnons and single-ion two-magnon bound states. In fact, the G mode 
absorbs most of the intensity and is thus far from extraneous. This mode is especially interesting in relation to the fact that the corresponding resonance 
line $\omega_G=g\mu_BH+D$ is an exact theoretical prediction both within the 3D model of Eq. (\ref{eq1}) and the 1D model of Eq. (\ref{eq16}); see Eqs. (\ref{eq10}) and 
(\ref{eq25}). 
\par 
Our numerical calculation also sheds light on the structure of the magnetic excitation spectrum in the intermediate phase $H_1<H<H_2$ where 
analytical results are practically absent. As is evident in Fig. \ref{fig4}, a tail of line G with strong intensity survives in the intermediate region even at 
low temperature where line G itself looses its intensity for $H>H_2$. Such a tail should thus be attributed to a high-frequency collective excitation 
that appears in the intermediate phase as a shadow of the single-ion two-magnon bound state, an issue that deserves further theoretical attention. 
\par Finally, the current calculation does not support the occurrence of a low-frequency Y structure suggested by Cox et al.\cite{cite6}, even though
 they also employ the 1D model of Eq. (\ref{eq16}) to calculate the susceptibility $\chi''(\omega)$. In fact, we find a V structure with rapidly decreasing 
intensity near the center of the intermediate phase. As such the V structure is expected to be especially vulnerable to small perturbations that are ever 
present in effective Heisenberg models. This may explain the deformation of the V into a U shape in the presence of a small Dzyaloshinskii-Moriya anisotropy \cite{cite5}. 

\section*{ACKNOWLEDGEMENTS}
The theoretical part of this work has been supported by the INT 238475 LOTHERM project. The experimental part was performed at the National High Magnetic Field 
Laboratory, Tallahassee, FL, which is supported by NSF Cooperative Agreement No. DMR-0654118, by the State of Florida, and by the DOE. S.Z. appreciates the support  of the  Deutsche 
Forschungsgemeinschaft and EuroMagNET II (EU contract No. 228043). 
We are grateful to Dr A.V Sizanov and Dr J. Oitmaa for useful correspondence concerning the results of Refs. \onlinecite{cite8} and \onlinecite{cite12}, respectively.


\begin{thebibliography}{47}
\expandafter\ifx\csname natexlab\endcsname\relax\def\natexlab#1{#1}\fi
\expandafter\ifx\csname bibnamefont\endcsname\relax
  \def\bibnamefont#1{#1}\fi
\expandafter\ifx\csname bibfnamefont\endcsname\relax
  \def\bibfnamefont#1{#1}\fi
\expandafter\ifx\csname citenamefont\endcsname\relax
  \def\citenamefont#1{#1}\fi
\expandafter\ifx\csname url\endcsname\relax
 \def\url#1{\texttt{#1}}\fi
\expandafter\ifx\csname urlprefix\endcsname\relax\def\urlprefix{URL }\fi
\providecommand{\bibinfo}[2]{#2}
\providecommand{\eprint}[2][]{\url{#2}}


\bibitem{cite1}
\bibinfo{author}{\bibfnamefont{V. S.}~\bibnamefont{Zapf,}} 
  \bibinfo{author}{\bibfnamefont{D.}~\bibnamefont{Zocco,}}
\bibinfo{author}{\bibfnamefont{B. R.}~\bibnamefont{Hansen,}}
\bibinfo{author}{\bibfnamefont{M.}~\bibnamefont{Jaime,}}
\bibinfo{author}{\bibfnamefont{N.}~\bibnamefont{Harrison,}}
\bibinfo{author}{\bibfnamefont{C. D.}~\bibnamefont{Batista,}}
\bibinfo{author}{\bibfnamefont{M.}~\bibnamefont{Kenzelmann,}}
\bibinfo{author}{\bibfnamefont{C.}~\bibnamefont{Niedermayer,}}
\bibinfo{author}{\bibfnamefont{A.}~\bibnamefont{Lacerda,}} and  
\bibinfo{author}{\bibfnamefont{A.}~\bibnamefont{Paduan-Filho}},
\bibinfo{journal}{Phys. Rev. Lett} \textbf{\bibinfo{volume}{96}},
  \bibinfo{pages}{077204} (\bibinfo{year}{2006})

\bibitem{cite2}
\bibinfo{author}{\bibfnamefont{S. A.}~\bibnamefont{Zvyagin,}} 
  \bibinfo{author}{\bibfnamefont{J.}~\bibnamefont{Wosnitza,}}
\bibinfo{author}{\bibfnamefont{C.D.}~\bibnamefont{Batista,}}
\bibinfo{author}{\bibfnamefont{M.}~\bibnamefont{Tsukamoto,}}
\bibinfo{author}{\bibfnamefont{N.}~\bibnamefont{Kawashima,}}
\bibinfo{author}{\bibfnamefont{J.}~\bibnamefont{Krzystek,}}
\bibinfo{author}{\bibfnamefont{V. S.}~\bibnamefont{Zapf,}}
\bibinfo{author}{\bibfnamefont{M.}~\bibnamefont{Jaime,}}
\bibinfo{author}{\bibfnamefont{N. F.}~\bibnamefont{Oliveira, Jr.,}} and
\bibinfo{author}{\bibfnamefont{A.}~\bibnamefont{Paduan-Filho}},
\bibinfo{journal}{Phys. Rev. Lett} \textbf{\bibinfo{volume}{98}},
  \bibinfo{pages}{047205} (\bibinfo{year}{2007})

\bibitem{cite3}
\bibinfo{author}{\bibfnamefont{S. A.}~\bibnamefont{Zvyagin,}} 
  \bibinfo{author}{\bibfnamefont{C. D.}~\bibnamefont{Batista,}}
\bibinfo{author}{\bibfnamefont{J.}~\bibnamefont{Krzystek,}}
\bibinfo{author}{\bibfnamefont{V. S.}~\bibnamefont{Zapf,}}
\bibinfo{author}{\bibfnamefont{M.}~\bibnamefont{Jaime,}}
\bibinfo{author}{\bibfnamefont{A.}~\bibnamefont{Paduan-Filho,}} and
\bibinfo{author}{\bibfnamefont{J.}~\bibnamefont{Wosnitza}} ,
\bibinfo{journal}{Physica B} \textbf{\bibinfo{volume}{403}},
  \bibinfo{pages}{1497} (\bibinfo{year}{2008})

\bibitem{cite4}
\bibinfo{author}{\bibfnamefont{N.}~\bibnamefont{Papanicolaou,}} 
  \bibinfo{author}{\bibfnamefont{A.}~\bibnamefont{Orend\'{a}\v{c}ov\'{a},}} and
\bibinfo{author}{\bibfnamefont{M.}~\bibnamefont{Orend\'{a}\v{c}}},
\bibinfo{journal}{Phys. Rev. B} \textbf{\bibinfo{volume}{56}},
  \bibinfo{pages}{8786} (\bibinfo{year}{1997})

\bibitem{cite5}
\bibinfo{author}{\bibfnamefont{S. A.}~\bibnamefont{Zvyagin,}} 
  \bibinfo{author}{\bibfnamefont{J.}~\bibnamefont{Wosnitza,}}
\bibinfo{author}{\bibfnamefont{A. K.}~\bibnamefont{Kolezhuk,}}
\bibinfo{author}{\bibfnamefont{V. S.}~\bibnamefont{Zapf,}}
\bibinfo{author}{\bibfnamefont{M.}~\bibnamefont{Jaime,}}
\bibinfo{author}{\bibfnamefont{A.}~\bibnamefont{Paduan-Filho,}}
\bibinfo{author}{\bibfnamefont{V. N.}~\bibnamefont{Glazkov,}} 
\bibinfo{author}{\bibfnamefont{S. S.}~\bibnamefont{Sosin,}} and
\bibinfo{author}{\bibfnamefont{A. I.}~\bibnamefont{Smirnov}},
\bibinfo{journal}{Phys. Rev. B} \textbf{\bibinfo{volume}{77}},
  \bibinfo{pages}{092413} (\bibinfo{year}{2008})

\bibitem{cite6}
\bibinfo{author}{\bibfnamefont{S.}~\bibnamefont{Cox,}} 
  \bibinfo{author}{\bibfnamefont{R. D.}~\bibnamefont{McDonald,}}
\bibinfo{author}{\bibfnamefont{M.}~\bibnamefont{Armanious,}}
\bibinfo{author}{\bibfnamefont{P.}~\bibnamefont{Sengupta,}} and
\bibinfo{author}{\bibfnamefont{A.}~\bibnamefont{Paduan-Filho}},
\bibinfo{journal}{Phys. Rev. Lett.} \textbf{\bibinfo{volume}{101}},
  \bibinfo{pages}{087602} (\bibinfo{year}{2008})

\bibitem{cite7}
\bibinfo{author}{\bibfnamefont{S. A.}~\bibnamefont{Zvyagin,}} 
\bibinfo{author}{\bibfnamefont{J.}~\bibnamefont{Krzystek,}} 
\bibinfo{author}{\bibfnamefont{P.H.M.}~\bibnamefont{van Loosderecht,}} 
\bibinfo{author}{\bibfnamefont{G.}~\bibnamefont{Dhalenne,}} and
\bibinfo{author}{\bibfnamefont{A.}~\bibnamefont{Revcolewschi,}} 
\bibinfo{journal}{Physica B} \textbf{\bibinfo{volume}{346-347}},
  \bibinfo{pages}{1} (\bibinfo{year}{2004})

\bibitem{cite8}
\bibinfo{author}{\bibfnamefont{A. V.}~\bibnamefont{Sizanov}} and
\bibinfo{author}{\bibfnamefont{A. V.}~\bibnamefont{Syromyatnikov,}}  
\bibinfo{journal}{arXiv:1105.1077v1 [cond-mat.str-el]} 

\bibitem{cite9}
\bibinfo{author}{\bibfnamefont{N.}~\bibnamefont{Papanicolaou}} and
\bibinfo{author}{\bibfnamefont{G. C.}~\bibnamefont{Psaltakis,}} 
\bibinfo{journal}{Phys. Rev. B} \textbf{\bibinfo{volume}{35}},
\bibinfo{pages}{342} (\bibinfo{year}{1987})

\bibitem{cite10}
\bibinfo{author}{\bibfnamefont{M.}~\bibnamefont{Wortis,}} 
\bibinfo{journal}{Phys. Rev.} \textbf{\bibinfo{volume}{132}},
\bibinfo{pages}{85} (\bibinfo{year}{1963})
 
\bibitem{cite11}
\bibinfo{author}{\bibfnamefont{N.}~\bibnamefont{Papanicolaou}} and 
\bibinfo{author}{\bibfnamefont{P. N.}~\bibnamefont{Spathis,}} 
\bibinfo{journal}{Phys. Rev. B} \textbf{\bibinfo{volume}{52}},
\bibinfo{pages}{16001} (\bibinfo{year}{1995})
 
\bibitem{cite12}
\bibinfo{author}{\bibfnamefont{A. F.}~\bibnamefont{Albuquerque,}} 
\bibinfo{author}{\bibfnamefont{C. J.}~\bibnamefont{Hamer,}} and
\bibinfo{author}{\bibfnamefont{J.}~\bibnamefont{Oitmaa}},
\bibinfo{journal}{Phys. Rev. B} \textbf{\bibinfo{volume}{79}},
\bibinfo{pages}{054412} (\bibinfo{year}{2009})

\bibitem{cite13}
\bibinfo{author}{\bibfnamefont{J.}~\bibnamefont{Oitmaa,}} 
\bibinfo{journal}{private communication.}

\bibitem{cite14}
\bibinfo{author}{\bibfnamefont{C. P.}~\bibnamefont{Slichter}},
  \emph{\bibinfo{booktitle}{Principles of Magnetic Resonance}}
  (\bibinfo{publisher}{Springer-Verlag, Berlin}, \bibinfo{year}{1978})

\bibitem{cite15}
\bibinfo{author}{\bibfnamefont{S. A.}~\bibnamefont{Zvyagin,}} 
\bibinfo{author}{\bibfnamefont{V. V.}~\bibnamefont{Eremenco,}}
\bibinfo{author}{\bibfnamefont{V. V.}~\bibnamefont{Pishko,}}
\bibinfo{author}{\bibfnamefont{A.}~\bibnamefont{Feher,}} 
\bibinfo{author}{\bibfnamefont{M.}~\bibnamefont{Orend\'{a}\v{c},}} and
\bibinfo{author}{\bibfnamefont{A.}~\bibnamefont{Orend\'{a}\v{c}ov\'{a},}}
\bibinfo{journal}{Low Temp. Phys.} \textbf{\bibinfo{volume}{21}},
\bibinfo{pages}{680} (\bibinfo{year}{1995})

\bibitem{cite16}
\bibinfo{author}{\bibfnamefont{S. A.}~\bibnamefont{Zvyagin,}} 
\bibinfo{author}{\bibfnamefont{T.}~\bibnamefont{Rieth,}}
\bibinfo{author}{\bibfnamefont{M.}~\bibnamefont{Sieling,}}
\bibinfo{author}{\bibfnamefont{S.}~\bibnamefont{Schmidt,}} and
\bibinfo{author}{\bibfnamefont{B.}~\bibnamefont{Luthi,}}
\bibinfo{journal}{Czech. J. Phys.} \textbf{\bibinfo{volume}{46}},
\bibinfo{pages}{1937} (\bibinfo{year}{1996})

\bibitem{cite18}
\bibinfo{author}{\bibfnamefont{A. K.}~\bibnamefont{Kolezhuk,}} and
\bibinfo{author}{\bibfnamefont{H.-J.}~\bibnamefont{Mikeska,}}
\bibinfo{journal}{Phys. Rev. B} \textbf{\bibinfo{volume}{65}},
\bibinfo{pages}{014413} (\bibinfo{year}{2001})

\end{thebibliography}
\end{document}